\documentclass[preprint,showpacs,preprintnumbers,superscriptaddress,nofootinbib]{revtex4}

\usepackage{graphicx}
\usepackage{dcolumn}
\usepackage{bm}
\usepackage{amsmath}
\usepackage{amssymb}
\usepackage{url}
\usepackage[normalem]{ulem}
\usepackage{xcolor}
\usepackage{listings} 
\newcommand{\beq}{\begin{equation}}
\newcommand{\eeq}{\end{equation}}
\newcommand{\bea}{\begin{eqnarray}}
\newcommand{\eea}{\end{eqnarray}}
\newcommand{\real}{{\sf I}\kern-.12em{\sf R}}
\newcommand{\comp}{{\sf I}\kern-.50em{\sf C}}
\newcommand{\unity}{{\sf I}\kern-.54em{\sf 1}}

\def\spose#1{\hbox to 0pt{#1\hss}}
\def\ltapprox{\mathrel{\spose{\lower 3pt\hbox{$\mathchar"218$}}
 \raise 2.0pt\hbox{$\mathchar"13C$}}}

\begin{document}

\title{Unintegrated gluon distribution from forward polarized
  $\rho$-electroproduction}
\author{Andr\`ee Dafne Bolognino}
\email{ad.bolognino@unical.it}
\affiliation{Dipartimento di Fisica dell'Universit\`a della Calabria \\
I-87036 Arcavacata di Rende, Cosenza, Italy}
\affiliation{INFN - Gruppo collegato di Cosenza, I-87036 Arcavacata di Rende,
  Cosenza, Italy}
\author{Francesco G. Celiberto}
\email{francescogiovanni.celiberto@unipv.it}
\affiliation{Instituto de F\'{\i}sica Te\'orica UAM/CSIC \\
  and Universidad Aut\'onoma de Madrid, Nicol\'as Cabrera 15, E-28049 Madrid,
  Spain}
\author{Dmitry Yu. Ivanov}
\email{d-ivanov@math.nsc.ru}
\affiliation{Sobolev Institute of Mathematics, 630090 Novosibirsk, Russia}
\affiliation{Novosibirsk State University, 630090 Novosibirsk, Russia}
\author{Alessandro Papa}
\email{alessandro.papa@fis.unical.it}
\affiliation{Dipartimento di Fisica dell'Universit\`a della Calabria \\
I-87036 Arcavacata di Rende, Cosenza, Italy}
\affiliation{INFN - Gruppo collegato di Cosenza, I-87036 Arcavacata di Rende,
Cosenza, Italy}

\date{\today}

\begin{abstract}
  We present here some arguments to support our suggestion that data on the
  helicity structure for the hard exclusive electroproduction of $\rho$ mesons
  at HERA (and in possible future high-energy electron-proton colliders)
  provide useful information to constrain the $\kappa$-shape of the unintegrated
  gluon distribution in the proton.
\end{abstract}

\maketitle

\section{Introduction}
\label{introd}

Our ability to find new Physics and understand the dynamics of strong
interactions at the LHC strongly relies on getting a more and more precise
 knowledge of  the structure of the proton.  In general, the  latter 
is encoded in different types of partonic distribution functions that enter
the factorization formalism for the description of the hard processes.
{\em Collinear factorization} is the most developed approach  to
calculate cross sections of inclusive reactions as a power expansion over the
 hard-scale  parameter. A prominent example here is the deep inelastic
scattering (DIS) of an electron off a proton. Its cross section, at the leading
order in the power expansion over the virtuality $Q^2$ of the exchanged photon
$\gamma^*$, is factorized as a convolution of a hard cross sections (calculable
in perturbation theory) with parton distribution functions (PDFs) of quarks
and gluons, $q_i(\zeta,\mu_F)$ and $g(\zeta,\mu_F)$, that depend on the
longitudinal momentum fraction of the proton carried by the parton, $\zeta$,
and on the factorization scale $\mu_F$, and obey DGLAP evolution
equations~\cite{DGLAP}. 
At the leading order (LO) of perturbation theory the variable $\zeta$ coincides
with the Bjorken variable $x=Q^2/(W^2+Q^2)$, where $W^2$ is the squared
center-of-mass energy of the $\gamma^* p$ system.  The collinear factorization
scheme can be also applied to the amplitudes of hard exclusive processes, where
the nonperturbative part is factorized in generalized parton
distributions~\cite{Collins:1996fb,Radyushkin:1997ki}.   
   
At high energy, $W\gg Q\gg \Lambda_{\rm QCD}$, the application of collinear
factorization is limited because the perturbative expansion includes in this
kinematics large logarithms of the energy that have to be resummed. Such a
resummation is incorporated in the {\em $\kappa$-factorization}\footnote{In this paper we use the expression ``$\kappa$-factorization'' to mean what elsewhere is known also as ``$k_T$-factorization''.}. The scattering
amplitudes are basically written as a convolution of the unintegrated gluon
distribution (UGD) in the proton with the impact factor (IF) that depends on
the  considered  process.
In the DIS case the $\gamma^* \to \gamma^*$   IF is calculated fully
in perturbation theory. The UGD is a nonperturbative quantity, function of
$x$ and $\kappa$, where the latter represents the gluon momentum transverse
to the direction of the proton and is the Fourier-conjugate variable of the
transverse separation $r_d$ of the color dipole into which the virtual
photon splits. Therefore small values of $r_d$ mean large values of $\kappa$
and {\it vice versa}. The UGD, in its original definition, obeys the
BFKL~\cite{BFKL} evolution equation in the $x$ variable.
Differently from collinear PDFs, the UGD is not well known and several types of 
models for it do exist, which lead to very different shapes in the
$(x,\kappa)$-plane (see, for instance, Refs.~\cite{small_x_WG,Angeles-Martinez:2015sea}).

The aim of this paper is to present our arguments that HERA data on
polarization observables in vector meson (VM) electroproduction can be
used to constrain the $\kappa$-dependence of the UGD in the HERA energy range.
In particular, we will focus our attention on the {\em ratio} of the two
dominant amplitudes for the polarized electroproduction of $\rho$ mesons,
{\it i.e.} the longitudinal VM production from longitudinally polarized
virtual photons and the transverse VM production from transversely polarized
virtual photons.

The H1 and ZEUS collaborations performed a complete
analysis~\cite{Aaron:2009xp,Chekanov:2007zr} of the spin density matrix elements
describing the hard exclusive light vector meson production, which can be
expressed in terms of helicity amplitudes for this process. 
The HERA data show distinctive features for both longitudinal and transverse VM
production: the same $W$- and $t$-dependence, that are  different 
from those seen in soft exclusive reactions (like VM photoproduction).
This  supports the
idea that the same physical mechanism, involving the scattering of a small
transverse size color dipole on the proton target, is at work for both helicity
amplitudes. Contrary to DIS case, the IFs for $\gamma^*\to \rm{VM}$ transitions
are not fully perturbative,  since  they include information about the VM
bound state. However, assuming the small size dipole dominance, one can
calculate the $\gamma^*\to \rm{VM}$ IFs unambiguously in collinear
factorization, as a convolution of the amplitudes of perturbative subprocesses
with VM distribution amplitudes (DAs) of twist-2 and
twist-3~\cite{Anikin:2009bf}. Such approach to helicity amplitudes of VM
electroproduction was used earlier in Ref.~\cite{Anikin:2011sa}, where a rather
simple model for UGD was adopted.      

Note that the $\kappa$-dependence of the IFs is different in the cases of
longitudinal and transverse polarizations and this poses a strong constraint
on the $\kappa$-dependence of the UGD in the HERA energy range. The main point
of our work will be to demonstrate, considering different models for UGD, that
the uncertainties of the theoretical description do not prevent us from some,
at least qualitative, conclusions about the $\kappa$-shape of the UGD. 

In this paper we concentrate on the $\kappa$-factorization method. The dipole
approach is based on similar physical ideas, but formulated not in $\kappa$-
but in the transverse coordinate space; this scheme is especially suitable to
account for nonlinear evolution and gluon saturation effects.  Interesting
developments are  the results of the papers~\cite{Besse:2012ia,Besse:2013muy},
where the helicity amplitudes of VM production were considered in the dipole
approach.

The paper is organized as follows: in Section~\ref{theory} we will present
the expressions for the amplitudes of interest here, discuss the
sources of theoretical uncertainties and sketch the main properties of
a few models for the UGD; in Section~\ref{analysis} we compare theoretical
predictions from the different models of UGD with HERA data; in
Section~\ref{discussion} we draw our conclusions and give some outlook.

\section{Theoretical setup}
\label{theory}

The H1 and ZEUS collaborations have provided extended analyses
of the helicity structure in the hard exclusive production of the $\rho$ meson
in $ep$ collisions through the subprocess
\begin{equation}
\label{process}
\gamma^*(\lambda_\gamma)p\rightarrow \rho (\lambda_\rho)p\,.
\end{equation}
Here $\lambda_\rho$ and $\lambda_\gamma$ represent the meson and photon
helicities, respectively, and can take the values 0 (longitudinal polarization)
and $\pm 1$ (transverse polarizations). The helicity amplitudes 
$T_{\lambda_\rho \lambda_\gamma}$ extracted at HERA~\cite{Aaron:2009xp} exhibit the
following hierarchy, that follows from the dominance of a  small-size 
dipole scattering mechanism, as discussed first in Ref.~\cite{Ivanov:1998gk}:
\begin{equation}
T_{00} \gg T_{11} \gg T_{10} \gg T_{01} \gg T_{-11}.
\end{equation}
The H1 and ZEUS collaborations have analyzed data in different ranges of $Q^2$
and $W$. In what follows we will refer only to the H1 ranges,
\begin{equation}
\begin{split}
2.5\,\text{\rm GeV$^2$} < Q^2 <60 \,\text{\rm GeV$^2$},\\
35\, \text{GeV} < W < 180\,\text{GeV},
\end{split}
\end{equation}
and will concentrate only on the dominant helicity ratio, $T_{11}/T_{00}$.

\subsection{Electroproduction of polarized $\rho$ mesons in the
$\kappa$-factorization}
\label{leptoproduction}

In the high-energy regime, $s\equiv W^2\gg Q^2\gg\Lambda_{\rm QCD}^2$, which
implies small $x=Q^2/W^2$, the forward helicity amplitude for the
$\rho$-meson electroproduction can be written, in $\kappa$-factorization, as
the convolution of the $\gamma^*\rightarrow \rho$ IF,
$\Phi^{\gamma^*(\lambda_\gamma)\rightarrow\rho(\lambda_\rho)}(\kappa^2,Q^2)$,
with the UGD, ${\cal F}(x,\kappa^2)$. Its expression reads
\begin{equation}
\label{amplitude}
  T_{\lambda_\rho\lambda_\gamma}(s,Q^2) = \frac{is}{(2\pi)^2}\int \dfrac{d^2\kappa}
  {(\kappa^2)^2}\Phi^{\gamma^*(\lambda_\gamma)\rightarrow\rho(\lambda_\rho)}(\kappa^2,Q^2)
  {\cal F}(x,\kappa^2),\quad \text{$x=\frac{Q^2}{s}$}\,.
\end{equation}

Defining $\alpha = \frac{\kappa^2}{Q^2}$ and $B =~2\pi \alpha_s
\frac{e}{\sqrt{2}}f_\rho$, the expression for the IFs takes the
following forms (see Ref.~\cite{Anikin:2009bf} for the derivation):

\begin{itemize}

\item longitudinal case
 
\begin{equation}
\label{Phi_LL}
\Phi_{\gamma_L\rightarrow\rho_L}(\kappa,Q;\mu^2) = 2 B\frac{\sqrt{N_c^2-1}}{Q\,N_c}
\int^{1}_{0}dy\, \varphi_1(y;\mu^2)\left(\frac{\alpha}{\alpha + y\bar{y}}\right)
\,,
\end{equation}
where $N_c$ denotes the number of colors and
$\varphi_1(y;\mu^2)$ is the twist-2 distribution amplitude (DA) which, up to
the second order in the expansion in Gegenbauer polynomials,
reads~\cite{Ball:1998sk}
\begin{equation}
\label{phi}
\varphi_1(y; \mu^2) = 6y\bar{y}\left(1+a_2(\mu^2)\frac{3}{2}
\left(5 (y-\bar{y})^2-1\right)\right)\,;
\end{equation}

\item transverse case

\[
\Phi_{\gamma_T\rightarrow\rho_T}(\alpha,Q;\mu^2)=
\dfrac{(\epsilon_{\gamma}\cdot\epsilon^{*}_{\rho}) \, 2 B m_{\rho}
  \sqrt{N_c^2-1}}{2 N_c Q^2}
\]
\begin{equation}
\times\left\{-\int^{1}_{0} dy \frac{\alpha (\alpha +2 y \bar{y})}{y\bar{y}
  (\alpha+y\bar{y})^2}\right. \left[(y-\bar{y})\varphi_1^T(y;\mu^2)
  + \varphi_A^T(y;\mu^2)\right]
\label{Phi_TT}
\end{equation}
\[
+\int^{1}_{0}dy_2\int^{y_2}_{0}dy_1 \frac{y_1\bar{y}_1\alpha}
     {\alpha+y_1\bar{y}_1}\left[\frac{2-N_c/C_F}{\alpha(y_1+\bar{y}_2)
         +y_1\bar{y}_2}
-\frac{N_c}{C_F}\frac{1}{y_2 \alpha+y_1(y_2-y_1)}\right]M(y_1,y_2;\mu^2)
\]
\[
\left.-\int^{1}_{0}dy_2\int^{y_2}_{0}dy_1 \left[\frac{2+N_c/C_F}{\bar{y}_1}
  +\frac{y_1}{\alpha+y_1\bar{y}_1}\right.\right.\left(\frac{(2-N_c/C_F)
    y_1\alpha}{\alpha(y_1+\bar{y}_2)+y_1\bar{y}_2}-2\right)
\]
\[
\left.\left.-\frac{N_c}{C_F}\frac{(y_2-y_1)\bar{y}_2}{\bar{y}_1}\frac{1}{\alpha\bar{y}_1+(y_2-y_1)\bar{y}_2}\right]\!S(y_1,y_2;\mu^2)\right\}\,,
\]
where $C_F=\frac{N_c^2-1}{2N_c}$, while the functions $M(y_1,y_2;\mu^2)$
and $S(y_1,y_2;\mu^2)$ are defined in Eqs.~(12)-(13) of
Ref.~\cite{Anikin:2011sa}
and are combinations of the twist-3 DAs $B(y_1,y_2;\mu^2)$ and
$D(y_1,y_2;\mu^2)$ (see Ref.~\cite{Ball:1998sk}), given by
\begin{align}
\label{BD}
B(y_1,y_2;\mu^2) & =-5040 y_1 \bar{y}_2 (y_1-\bar{y}_2) (y_2-y_1) \notag\\
D(y_1,y_2;\mu^2) & =-360 y_1\bar{y}_2(y_2-y_1)
\left(1+\frac{\omega^{A}_{\{1,0\}}(\mu^2)}{2}\left(7\left(y_2-y_1\right)-3\right)
\right)\,.
\end{align}

\end{itemize}

In Eqs.~\eqref{phi} and~\eqref{BD} the functional dependence of
$a_2$, $\omega^{A}_{\{1,0\}}$, $\zeta^{A}_{3\rho}$, and $\zeta^{V}_{3\rho}$ on the
factorization scale $\mu^2$ can be determined from the corresponding known
evolution equations~\cite{Ball:1998sk}, using some suitable initial condition
at a scale $\mu_0$. 

The DAs $\varphi^T_1(y;\mu^2)$ and $\varphi^T_A(y;\mu^2)$ in Eq.~\eqref{Phi_TT}
encompass both genuine twist-3 and Wandzura-Wilczek~(WW)
contributions~\cite{Ball:1998sk, Anikin:2011sa}. The former are related to
$B(y_1,y_2;\mu^2)$ and $D(y_1,y_2;\mu^2)$; the latter are those obtained
in the approximation in which $B(y_1,y_2;\mu^2)=D(y_1,y_2;\mu^2)=0$, and in
this case read\footnote{For asymptotic form of the twist-2 DA,
  $\varphi_1(y)=\varphi_1^{\rm{as}}(y)=6y\bar y$, these equations give
  $\varphi^{T\;WW,\;\rm{as}}_A(y)=-3/2y\bar y$ and $\varphi^{T\;WW,\;\rm{as}}_1(y)=
  -3/2 y\bar y (2y -1)$.}
\[
\label{WWT}
\varphi^{T\;WW}_A(y;\mu^2)= \frac{1}{2}\left[-\bar y
\int_0^{y}\,dv \frac{\varphi_1(v;\mu^2)}{\bar v} -
y \int_{y}^1\,dv \frac{\varphi_1(v;\mu^2)}{v}   \right]\;,
\]
\begin{equation}
\varphi^{T\;WW}_1(y;\mu^2)= \frac{1}{2}\left[
-\bar y \int_0^{y}\,dv \frac{\varphi_1(v;\mu^2)}{\bar v} +
y \int_{y}^1\,dv \frac{\varphi_1(v;\mu^2)}{v}   \right]\;.
\end{equation}

\subsection{Theoretical uncertainties and approximations}
\label{uncertainties}

There are four sources of uncertainty and/or approximation in our analysis,
as based on the above expressions for the helicity amplitudes.

(i) The $\gamma^*\to V$ IF is a function of $Q^2$ and $\kappa$ which is not
fully perturbative and includes also physics of large distances. Here we use
collinear factorization to express the IF as a convolution -- integration over
longitudinal momentum fraction --  of the nonperturbative twist-2 and -3
DAs and a perturbative hard part.

In the region of large $\kappa$, $\kappa\sim Q$, which corresponds to
the range of small dipole sizes, the IFs for the production of both the
longitudinally and transversely polarized meson are well described in our
collinear factorization scheme. The neglected contributions are relatively
suppressed as powers of $\Lambda_{\rm QCD}/Q$ and are therefore neglected.

The region of small $\kappa$, $\kappa\ll Q$, is also present in our
$\kappa$-factorization formulas and corresponds to the range of larger dipole
sizes $r_d$. Can we calculate also in this case our IFs as convolution of
the perturbative hard part with the meson DAs?

The situation here is different in the cases of longitudinal and transverse
polarizations. In the longitudinal case, we have, in fact, small $r_d$
dominance in the region of all $\kappa$. Note that,  as  calculated in our
scheme, the longitudinal IF divided by $\kappa^2$ is finite for $\kappa \to 0$, and, therefore, the  neglected contributions for our longitudinal IF are
suppressed by powers of $\Lambda_{\rm QCD}/Q$ in the region of all $\kappa$.
In relation with that, we note here that the longitudinal VM electroproduction
can be described not only in $\kappa$-, but also fully in QCD collinear
factorization (in terms of generalized parton distributions).

For the transverse polarization the situation is different, since
the transverse IF divided by $\kappa^2$ behaves like $\log(\kappa^2/Q^2)$,
which means that the collinear limit, $\kappa\to 0$, is not safe and one
cannot describe the transverse VM electroproduction fully in QCD collinear
factorization. One can easily trace that this behavior 
$\sim \log(\kappa^2/Q^2)$  for  $\kappa \to 0$ of the transverse IF appears
due to the  integration over the  longitudinal fraction $z$ and its
logarithmic divergence near the end points $z\to 0, 1$.
The light-cone wave function of the virtual photon, that controls the hard part
of the IF, has a scale $Q^2 r_d^2 z (1-z)$. This means that, at $\kappa \to 0$,
when the endpoint region of the $z$-integration is important, large values of
$Q$ do not mean automatically that the small-$r_d$ region is dominant, but
instead both large and small $r_d$ contribute. Therefore we do not control
the accuracy of our IF calculation for $\kappa \to 0$ in the transverse
polarization case.

However, in $\kappa$-factorization the small-$\kappa$ region is only a corner
of the $\kappa$ integration domain and the importance of this corner is a
matter of investigation. On the experimental side we do not see indications
that large $r_d$'s, and therefore small $\kappa$'s, are dominant; indeed,
HERA data show similar $t$- and $W$-dependence for the both longitudinal and
transverse helicity amplitudes. In our phenomenological analysis we will check
the importance of the region of small $\kappa$'s by studying the dependence of
our predictions on the $\kappa$ lower cut value.

(ii) Another source of uncertainty comes from the adopted form of the 
light-cone DAs.

We considered, for the sake of simplicity, the so called {\em asymptotic}
choice for the twist-2 DA given in Eq.~\eqref{phi}, corresponding
to fixing $a_2(\mu^2)=0$. The impact of this approximation was estimated by
letting $a_2(\mu_0^2)$ take a non-zero value as large as 0.6 at
$\mu_0^2=1$~GeV$^2$ in the analysis with one specific model for UGD.

We used typically twist-3 DAs in the Wandzura-Wilczek approximation, but
considered in one case the effect of the inclusion of the genuine twist-3
contribution to check the validity of this approximation.

(iii) We calculate the {\em forward} amplitudes for both longitudinal and
transverse case. The experimental analysis showed that the $t$-dependence is
similar for the two helicity amplitudes, the measured values of the slope
parameter have, within errors, the same values for the both polarizations cases.
Therefore in $T_{11}/T_{00}$ ratio  considered here  the $t$-dependence
drops. 

(iv) The expression in Eq.~\eqref{amplitude} represents, as a matter of fact,
the {\em imaginary part} of the amplitude and not the full amplitude. The real
parts of the amplitudes at high energy are smaller: they are suppressed in
comparison to the imaginary parts by the factor $\sim 1/\log s$, and they are
related to the latter by dispersion relations. Here again we appeal to the
results of the experimental analysis, that showed similar $W$-dependence for
both helicity amplitudes, that means the effective cancellation of the
 contribution  from the real parts of the amplitudes in the ratio
$T_{11}/T_{00}$.

\subsection{Models of Unintegrated Gluon Distribution}
\label{models}

In this work we have considered a selection of six models of UGD, without
pretension to exhaustive coverage, but with the aim of comparing
(sometimes radically) different approaches. We refer the reader to
the original papers for details on the derivation of each model and limit
ourselves to presenting here just the functional form ${\cal F}(x,\kappa^2)$
of the UGD as we implemented it in the numerical analysis.

\subsubsection{An $x$-independent model (ABIPSW)}

The simplest UGD model is $x$-independent and merely coincides with
the proton impact factor~\cite{Anikin:2011sa}:
\begin{equation}
  {\cal F}(x,\kappa^2)= \frac{A}{(2\pi)^2\,M^2}
  \left[\frac{\kappa^2}{M^2+\kappa^2}\right]\,,
\end{equation}
where $M$ corresponds to the non-perturbative hadronic scale. The constant $A$
is unessential since we are going to consider the ratio $T_{11}/T_{00}$.

\subsubsection{Gluon momentum derivative}

This UGD is given by
\begin{equation}
\label{xgluon}
{\cal F}(x, \kappa^2) = \frac{dxg(x, \kappa^2)}{d\ln \kappa^2}
\end{equation}
and encompasses the collinear gluon density $g(x, \mu_F^2)$, taken at
$\mu_F^2=\kappa^2$. It is based on the obvious requirement that, when
integrated over $\kappa^2$ up to some factorization scale, the UGD must
give the collinear gluon density. We have employed the CT14
parametrization~\cite{Dulat:2015mca}, using the appropriate cutoff
$\kappa_{\rm min} = 0.3$~GeV (see Section~\ref{tools} for further details).

\subsubsection{Ivanov--Nikolaev' (IN) UGD: a soft-hard model}

The UGD proposed in Ref.~\cite{Ivanov:2000cm} is developed with the purpose
of probing different regions of the transverse momentum. In the large-$\kappa$
region, DGLAP parametrizations for $g(x, \kappa^2)$ are employed. Moreover,
for the extrapolation of the hard gluon densities to small $\kappa^2$, an
Ansatz is made~\cite{Nikolaev:1994cd}, which describes the color gauge
invariance constraints on the radiation of soft gluons by color singlet
targets. The gluon density at small $\kappa^2$ is supplemented by a
non-perturbative soft component, according to the color-dipole
phenomenology.

This model of UGD has the following form:
\begin{equation}
{\cal F}(x,\kappa^2)= {\cal F}^{(B)}_\text{soft}(x,\kappa^2) 
{\kappa_{s}^2 \over 
	\kappa^2 +\kappa_{s}^2} + {\cal F}_\text{hard}(x,\kappa^2) 
{\kappa^2 \over 
	\kappa^2 +\kappa_{h}^2}\,,
\label{eq:4.7}
\end{equation}
where $\kappa_{s}^2 = 3$ GeV$^2$ and $\kappa_{h}^2 = [1 + 0.047\log^2(1/x)]^{1/2}$.

The soft term reads
\begin{equation}
\label{softterm}
{\cal F}^{(B)}_\text{soft}(x,\kappa^2) = a_\text{soft}
C_{F} N_{c} {\alpha_{s}(\kappa^2) \over \pi} \left( {\kappa^2 \over 
	\kappa^2 +\mu_\text{soft}^{2}}\right)^2 V_{\rm N}(\kappa)\,,
\end{equation}
where $C_{F} = \dfrac{N_{c}^2 -1}{2N_{c}}$ and $\mu_\text{soft} = 0.1$ GeV. The
parameter $a_\text{soft} = 2$ gives a measure of how important is the soft part
compared to the hard one. On the other hand, the hard component reads
\begin{equation}
\label{hardterm}
{\cal F}_\text{hard}(x,\kappa^2)= 
{\cal F}^{(B)}_{\text{pt}}(\kappa^2){{\cal F}_{\text{pt}}(x,Q_{c}^{2})
 	\over {\cal F}_{\text{pt}}^{(B)}(Q_{c}^{2})}
\theta(Q_{c}^{2}-\kappa^{2}) +{\cal F}_{\text{pt}}(x,\kappa^2)
\theta(\kappa^{2}-Q_{c}^{2})\,,
\end{equation}	
where ${\cal F}_{\text{pt}}(x, \kappa^2)$ is related to the standard gluon parton
distribution as in Eq.~\eqref{xgluon} and $Q_{c}^2 = 3.26$ GeV$^2$ 
(see Section~\ref{tools} for further details).
We refer to Ref.~\cite{Ivanov:2000cm} for the expressions of the vertex
function $V_{\rm N}(\kappa)$ and of ${\cal F}^{(B)}_{\text{pt}}(\kappa^2)$.
Another relevant feature of this model is given by the choice of the coupling
constant. In this regard, the infrared freezing of strong coupling at leading
order (LO) is imposed by fixing $\Lambda_\text{QCD} = 200$ MeV:
\begin{equation}
\label{frozen}
\alpha_s(\mu^2) = \text{min} \left\{0.82, \, \frac{4 \pi}{\beta_0
  \log \left(\frac{\mu^2}{\Lambda^2_\text{QCD}}\right)}\right\}.
\end{equation}

We stress that this model was successfully tested on the {\em unpolarized}
electroproduction of VMs at HERA.

\subsubsection{Hentschinski--Sabio Vera--Salas' (HSS) model}

This model, originally used in the study of DIS structure
functions~\cite{Hentschinski:2012kr}, takes the form of a convolution between
the BFKL gluon Green's function and a LO proton impact factor. It has been
employed in the description of single-bottom quark production at LHC
in Ref.~\cite{Chachamis:2015ona} and to investigate the photoproduction of
$J/\Psi$ and $\Upsilon$ in Ref.~\cite{Bautista:2016xnp}. We implemented
the formula given in Ref.~\cite{Chachamis:2015ona} (up to a $\kappa^2$ overall
factor needed to match our definition), which reads
\begin{equation}
\label{HentsUGD}
 {\cal F}(x, \kappa^2, M_h) = \int_{-\infty}^{\infty}
  \frac{d\nu}{2\pi^2}\ {\cal C} \  \frac{\Gamma(\delta - i\nu -\frac{1}{2})}
      {\Gamma(\delta)}\ \left(\frac{1}{x}\right)^{\chi\left(\frac{1}{2}+i\nu\right)}
      \left(\frac{\kappa^2}{Q^2_0}\right)^{\frac{1}{2}+i\nu}
\end{equation}
\[
\times \left\{ 1 +\frac{\bar{\alpha}^2_s \beta_0 \chi_0\left(\frac{1}{2}
+i\nu\right)}{8 N_c}\log\left(\frac{1}{x}\right)
\left[-\psi\left(\delta-\frac{1}{2} - i\nu\right)
-\log\frac{\kappa^2}{M_h^2}\right]\right\}\,,
\]
where $\beta_0=\frac{11 N_c-2 N_f}{3}$, with $N_f$ the number of
active quarks (put equal to four in the following),
$\bar{\alpha}_s = \dfrac{\alpha_s\left(\mu^2\right) N_c}{\pi}$,
with $\mu^2 = Q_0 M_h$, and $\chi_0(\frac{1}{2} + i\nu)\equiv \chi_0(\gamma)
= 2\psi(1) - \psi(\gamma) - \psi(1-\gamma)$ is  (up to the factor $\bar\alpha_s$) the LO eigenvalue of the BFKL
kernel, with $\psi(\gamma)$ the logarithmic derivative of Euler Gamma
function. Here, $M_h$ plays the role of the hard scale which can be identified
with the photon virtuality, $\sqrt{Q^2}$.
In Eq.~\eqref{HentsUGD}, $\chi(\gamma)$ (with $\gamma = \frac{1}{2} + i\nu$)
is the NLO eigenvalue of the BFKL kernel, collinearly improved and BLM
optimized; it reads
\begin{equation}
  \chi(\gamma) = \bar{\alpha}_s\chi_0(\gamma)+\bar{\alpha}^2_s\chi_1(\gamma)
  -\frac{1}{2}\bar{\alpha}^2_s\chi^\prime_0(\gamma)\,\chi_0(\gamma)
  + \chi_{RG}(\bar{\alpha}_s, \gamma)\,,
\end{equation}
with $\chi_1(\gamma)$ and $\chi_{RG}(\bar{\alpha}_s, \gamma)$ given in
Section~2 of Ref.~\cite{Chachamis:2015ona}.

This UGD model is characterized by a peculiar parametrization for the proton
impact factor, whose expression is
\begin{equation}
  \Phi_p(q, Q^2_0) = \frac{{\cal C}}{2\pi \Gamma(\delta)}
  \left(\frac{q^2}{Q^2_0}\right)^\delta e^{-\frac{q^2}{Q^2_0}},
\end{equation}
which depends on three parameters $Q_0$, $\delta$ and ${\cal C}$ which
were fitted to the combined HERA data for the $F_2(x)$ proton structure
function. We adopted here the so called
{\em kinematically improved} values (see Section~\ref{tools} for further
details) and given by
\begin{equation}
\label{ki}
Q_0 = 0.28\,\text{GeV}, \qquad \delta = 6.5, \qquad {\cal C} = 2.35 \;.
\end{equation}

\subsubsection{Golec-Biernat--W{\"u}sthoff' (GBW) UGD}

This UGD parametrization derives from the effective dipole cross section
$\hat{\sigma}(x,r)$ for the scattering of a $q\bar{q}$ pair off a
nucleon~\cite{GolecBiernat:1998js},
\begin{equation}
  \hat{\sigma}(x, r^2) = \sigma_0 \left\{1-\exp\left(-\frac{r^2}{4R^2_0(x)}
  \right)\right\}\,,
\end{equation}
through a reverse Fourier transform of the expression 
\begin{equation}
  \sigma_0 \left\{1-\exp\left(-\frac{r^2}{4R^2_0(x)}
  \right)\right\}=\int \frac{d^2\kappa}{\kappa^4} {\cal F}(x,\kappa^2)
  \left(1-\exp(i \vec{\kappa}\cdot\vec{r})\right)\left(1-\exp(-i \vec{\kappa}
  \cdot\vec{r})\right)\,,
\end{equation}
\begin{equation}
  {\cal F}(x,\kappa^2)= \kappa^4 \sigma_0 \frac{R^2_0(x)}{2\pi}
  e^{-\kappa^2 R^2_0(x)}\,,
\end{equation}
with 
\begin{equation}
R^2_0(x) = \frac{1}{{\rm GeV}^2} \left(\frac{x}{x_0}\right)^{\lambda_p}
\end{equation}
 and  the following values
\begin{equation}
\sigma_0 = 23.03\,\text{mb}, \qquad \lambda_p = 0.288, \qquad x_0 = 3.04 \cdot 10^{-4}\,.
\end{equation}
The normalization $\sigma_0$ and the parameters $x_0$ and $\lambda_p > 0$ of
$R^2_0(x)$ have been determined by a global fit to $F_2(x)$ in the
region $x < 0.01$.

\subsubsection{Watt--Martin--Ryskin' (WMR) model}

The UGD introduced in Ref.~\cite{Watt:2003mx} reads
\[
{\cal F}(x,\kappa^2,\mu^2) = T_g(\kappa^2,\mu^2)\,\frac{\alpha_s(\kappa^2)}
{2\pi}\,\int_x^1\!dz\;\left[\sum_q P_{gq}(z)\,\frac{x}{z}q\left(\frac{x}{z},
  \kappa^2\right) + \right.\nonumber
\]
\begin{equation}
  \label{WMR_UGD}
  \left. \hspace{6.5cm} P_{gg}(z)\,\frac{x}{z}g\left(\frac{x}{z},\kappa^2\right)\,\Theta\left(\frac{\mu}{\mu+\kappa}-z\right)\,\right]\,,
\end{equation}
where the term
\begin{equation}
\label{WMR_Tg}
T_g(\kappa^2,\mu^2) = \exp\left(-\int_{\kappa^2}^{\mu^2}\!d\kappa_t^2\,
\frac{\alpha_s(\kappa_t^2)}{2\pi}\,\left( \int_{z^\prime_{{\rm min}}}^{z^\prime_{{\rm max}}}
\!dz^\prime\;z^\prime \,P_{gg}(z^\prime ) + N_f\,\int_0^1\!dz^\prime\,P_{qg}(z^\prime)
\right)\right)\,,
\end{equation}
gives the probability of evolving from the scale $\kappa$ to the
scale $\mu$ without parton emission. Here $z^\prime_{\mathrm{max}}\equiv
1-z^\prime_{\mathrm{min}}=\mu/(\mu+\kappa_t)$; $N_f$ is the number of active quarks.
This UGD model depends on an extra-scale $\mu$, which we fixed at $Q$.
The splitting functions $P_{qg}(z)$ and $P_{gg}(z)$ are given by
\[
P_{qg}(z) = T_R\,[z^2 + (1-z)^2]\;,
\]
\[
P_{gg}(z) = 2\,C_A \left[\dfrac{1}{(1-z)_+} + \dfrac{1}{z}- 2 +z(1-z)\right]
+ \left(\frac{11}{6}C_A - \frac{N_f}{3}\right) \delta(1 -z)\;,
\]
with the plus-prescription defined as
\begin{equation}
\label{pluspre}
\int_{a}^{1} dz \frac{F(z)}{(1-z)_+} = \int_{a}^{1} dz \frac{F(z) - F(1)}{(1-z)}
- \int_{0}^{a} dz \frac{F(1)}{(1 -z)}\,.
\end{equation}

\section{Numerical analysis}
\label{analysis}

In this Section we present our results for the helicity-amplitude ratio
$T_{11}/T_{00}$, as obtained with the six UGD models presented above, and
compare them with HERA data.

\begin{figure}[tb]
\centering

\includegraphics[scale=0.60,clip]{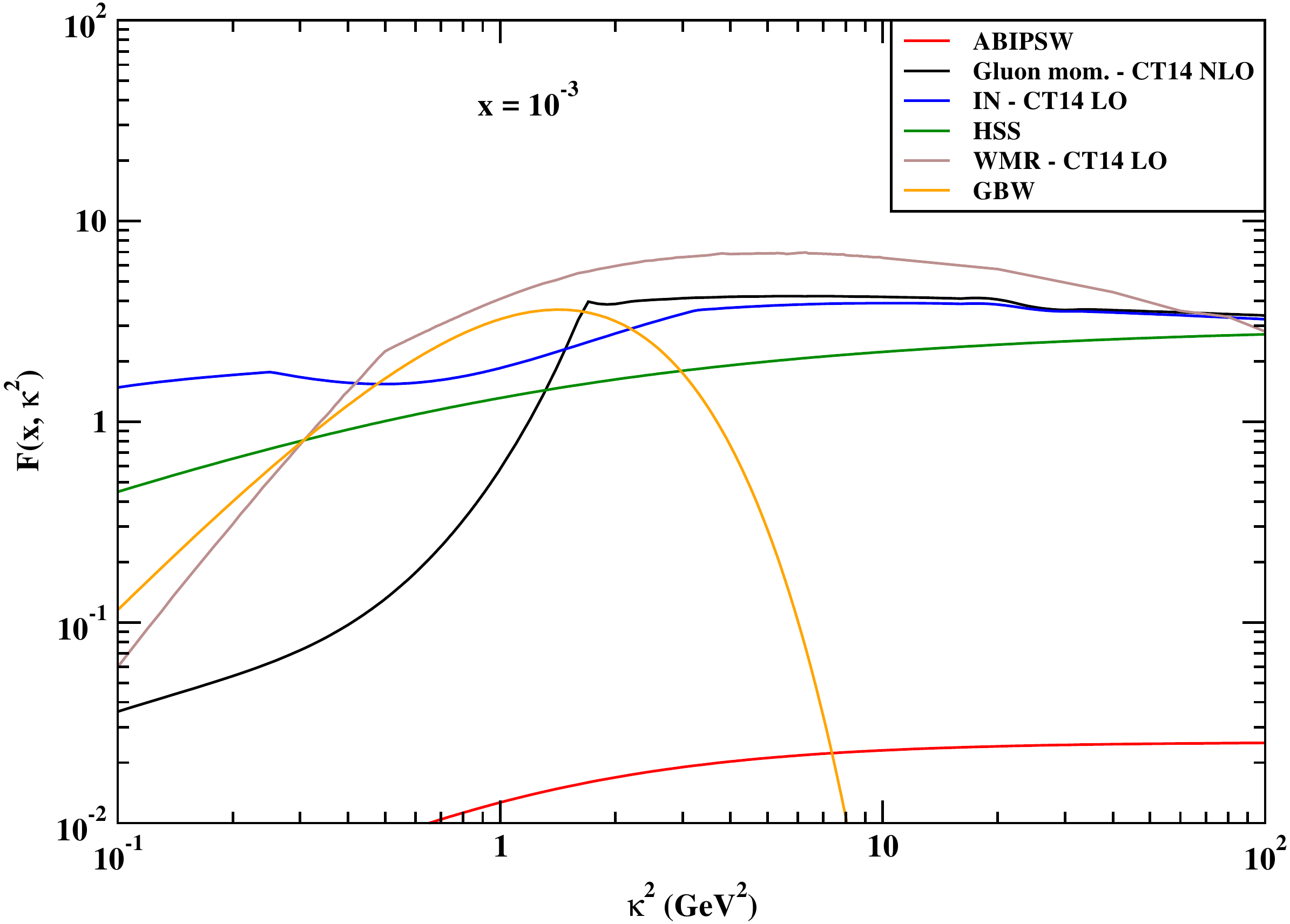}

\includegraphics[scale=0.60,clip]{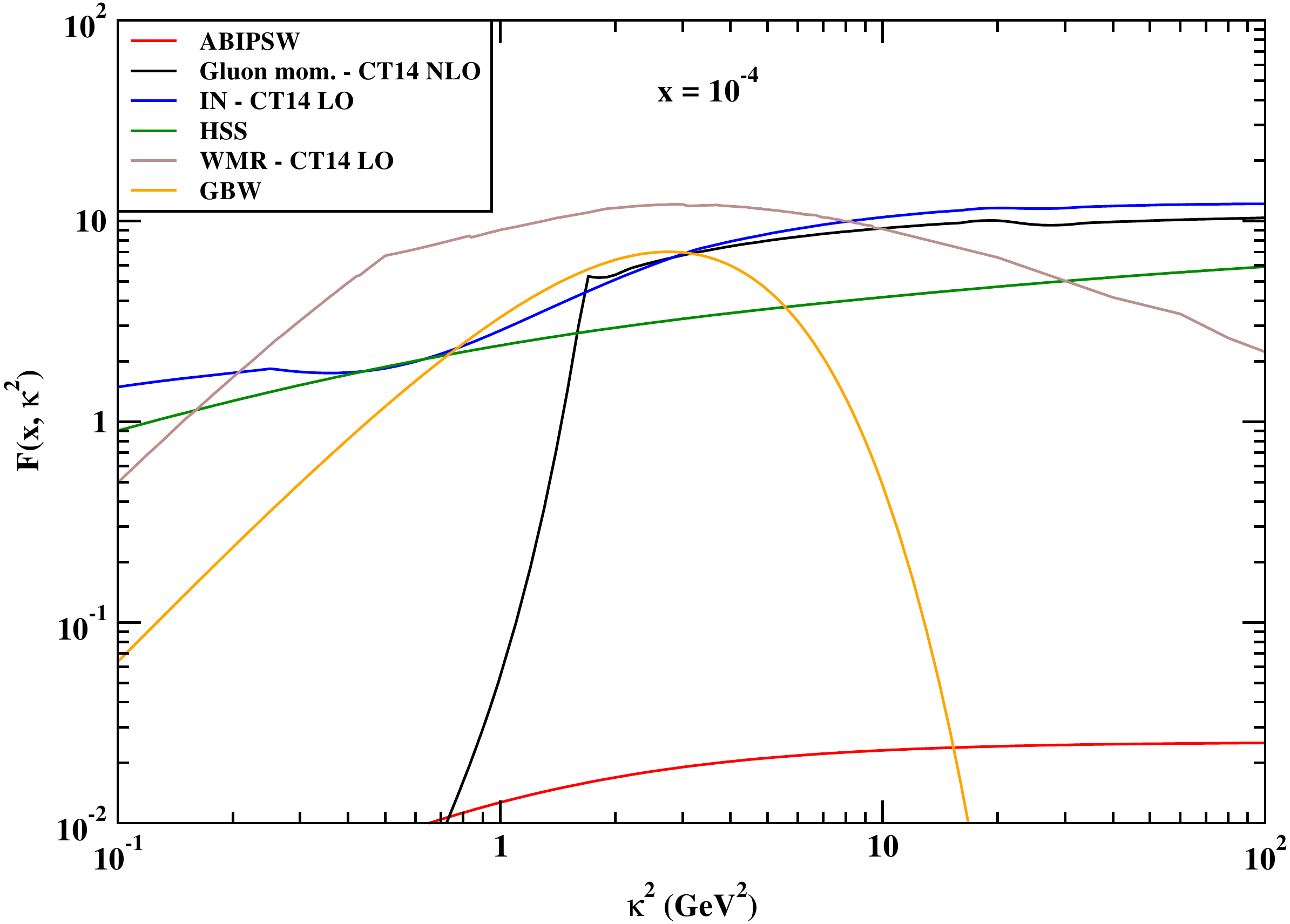}

\caption{$\kappa^2$-dependence of all UGD models for $x = 10^{-3}$ and $10^{-4}$.}
\label{fig:UGDs_vs_k2}
\end{figure}

We preliminarily present a plot, Fig.~\ref{fig:UGDs_vs_k2}, with the
$\kappa^2$-dependence of all the considered UGD models, for two different
values of $x$. The plot clearly exhibits the marked difference in the
$\kappa^2$-shape of the six UGDs.

\begin{figure}[tb]
\centering

\includegraphics[scale=0.60,clip]{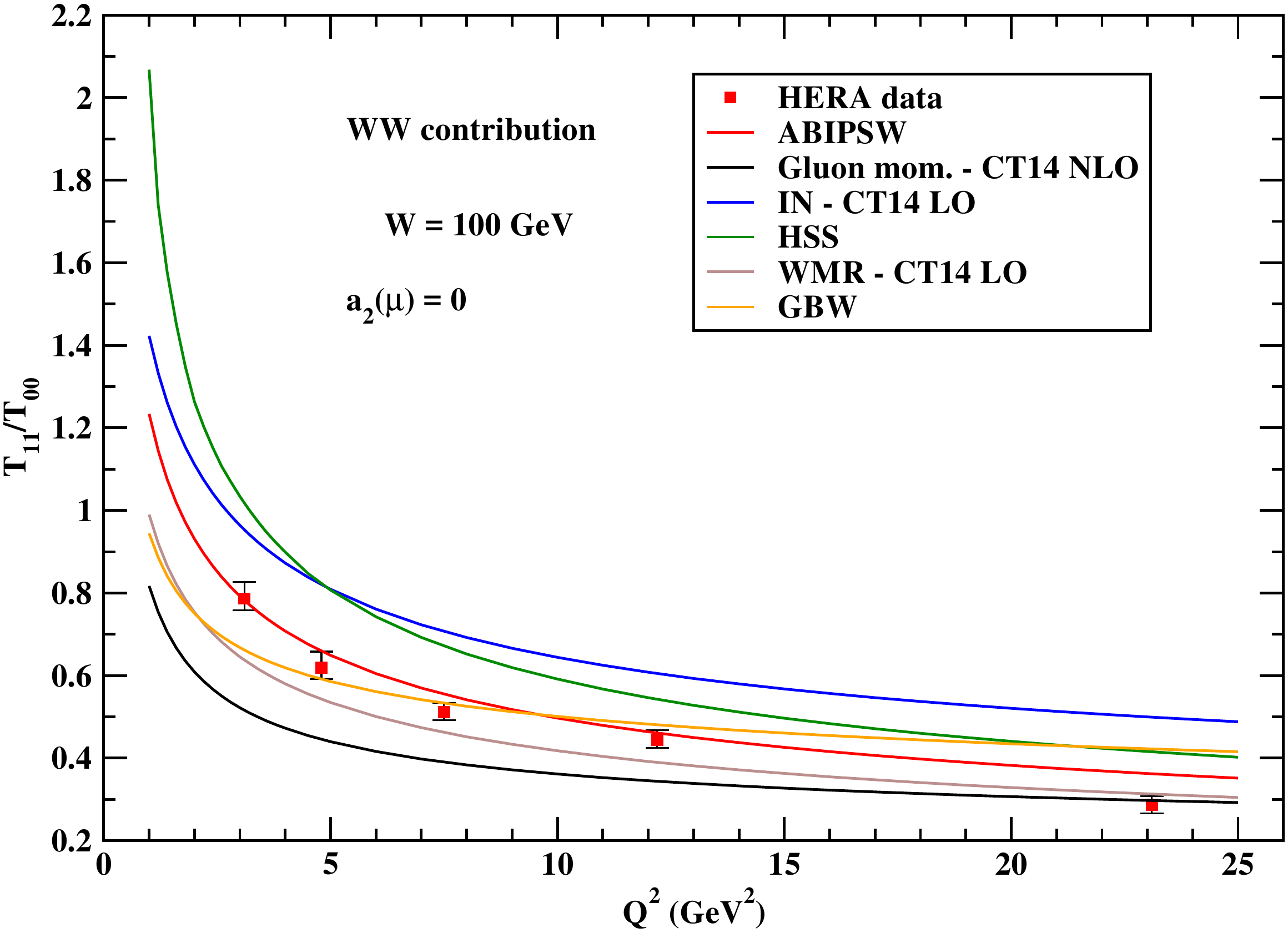}

\caption{$Q^2$-dependence of the helicity-amplitude ratio $T_{11}/T_{00}$ for
all the considered UGD models at $W = 100$ GeV. In the twist-2 DA we have
put $a_2(\mu_0 = 1 \mbox{ GeV}) = 0$ and the $T_{11}$ amplitude has
been calculated in the WW approximation.}
\label{fig:ratio_all}
\end{figure}

In Fig.~\ref{fig:ratio_all} we compare the $Q^2$-dependence of $T_{11}/T_{00}$
for all six models at $W = 100$~GeV, together with the experimental result.
We used here the asymptotic twist-2 DA ($a_2(\mu^2)=0$) and the WW approximation
for twist-3 contributions. Theoretical results are spread over a large interval,
thus supporting our claim that the observable $T_{11}/T_{00}$ is potentially
able to strongly constrain the $\kappa$-dependence of the UGD. None of the
models is able to reproduce data over the entire $Q^2$ range; the
$x$-independent ABIPSW model and the GBW model seem to better catch the
intermediate-$Q^2$ behavior of data.

\begin{figure}[tb]
\centering

\includegraphics[scale=0.60,clip]{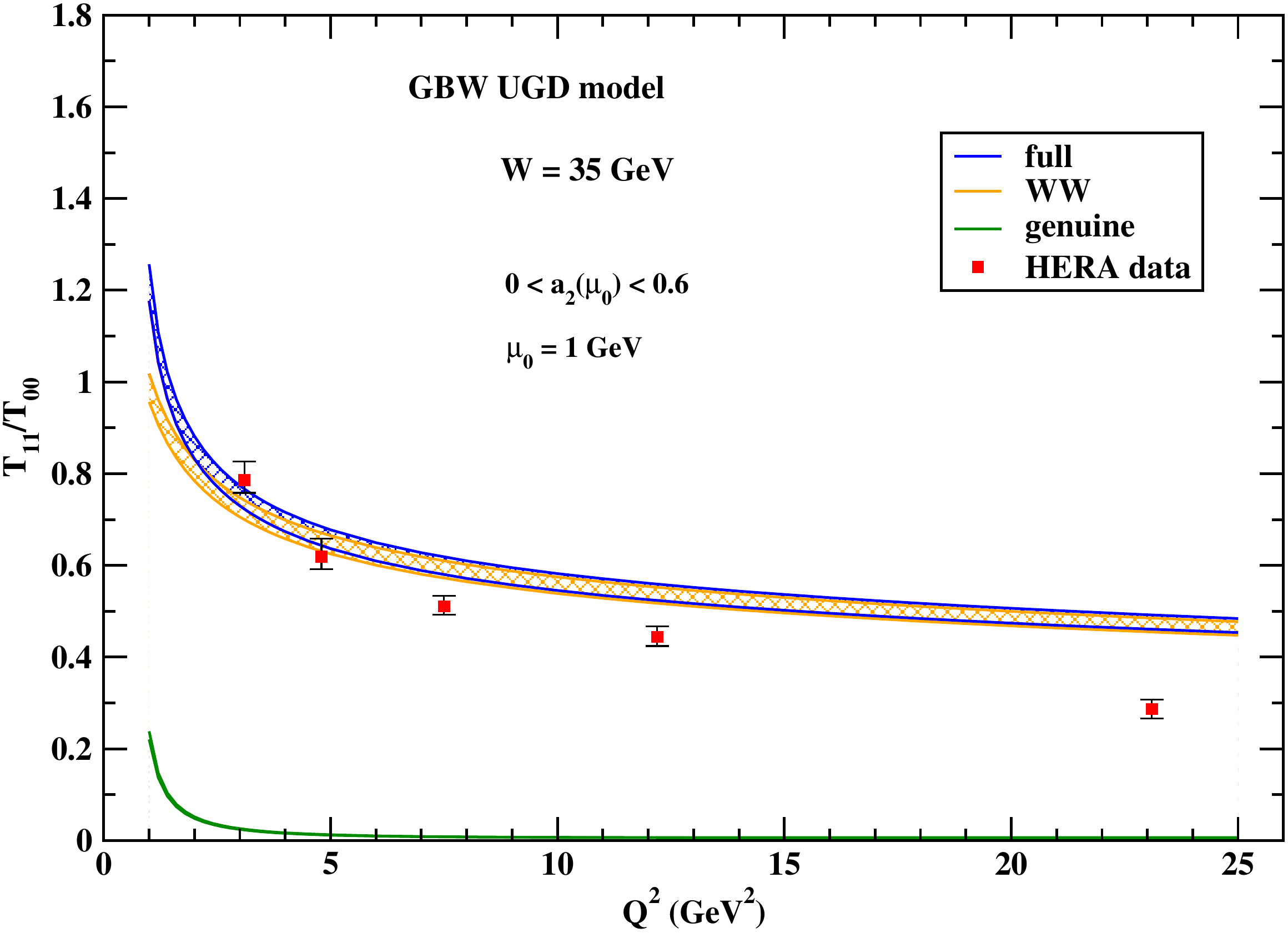}

\includegraphics[scale=0.60,clip]{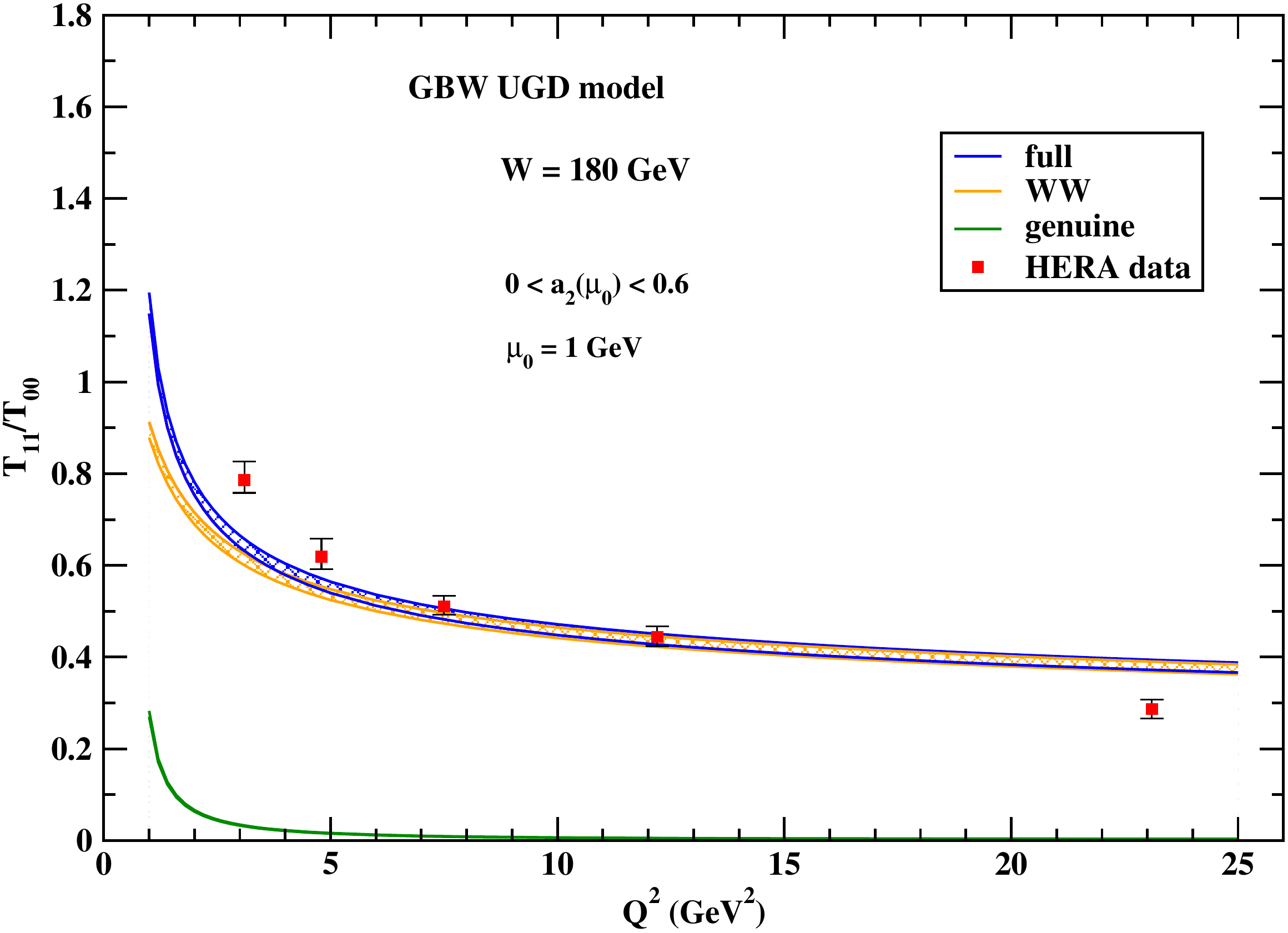}

\caption{$Q^2$-dependence of the helicity-amplitude ratio $T_{11}/T_{00}$ for
the GBW UGD model at $W = 35$ (top) and 180~GeV (bottom). The full, WW and
genuine contributions are shown. The shaded bands give the effect of
varying $a_2(\mu_0 = 1 \mbox{ GeV})$ between 0. and 0.6.}
\label{fig:ratio_GBW_evolved}
\end{figure}

To gauge the impact of the approximation made in the DAs, we calculated
the $T_{11}/T_{00}$ ratio with the GBW model, at $W = 35$ and 180~GeV,
by varying $a_2(\mu_0=1 \ {\rm GeV})$ in the range 0. to 0.6 and properly
taking into account its evolution. Moreover, for the same UGD model, we
relaxed the WW approximation in $T_{11}$ and considered also the genuine
twist-3 contribution. All that is summarized in
Fig.~\ref{fig:ratio_GBW_evolved}, which indicates that the approximations
made are quite reliable.

\begin{figure}[tb]
\centering

\includegraphics[scale=0.60,clip]{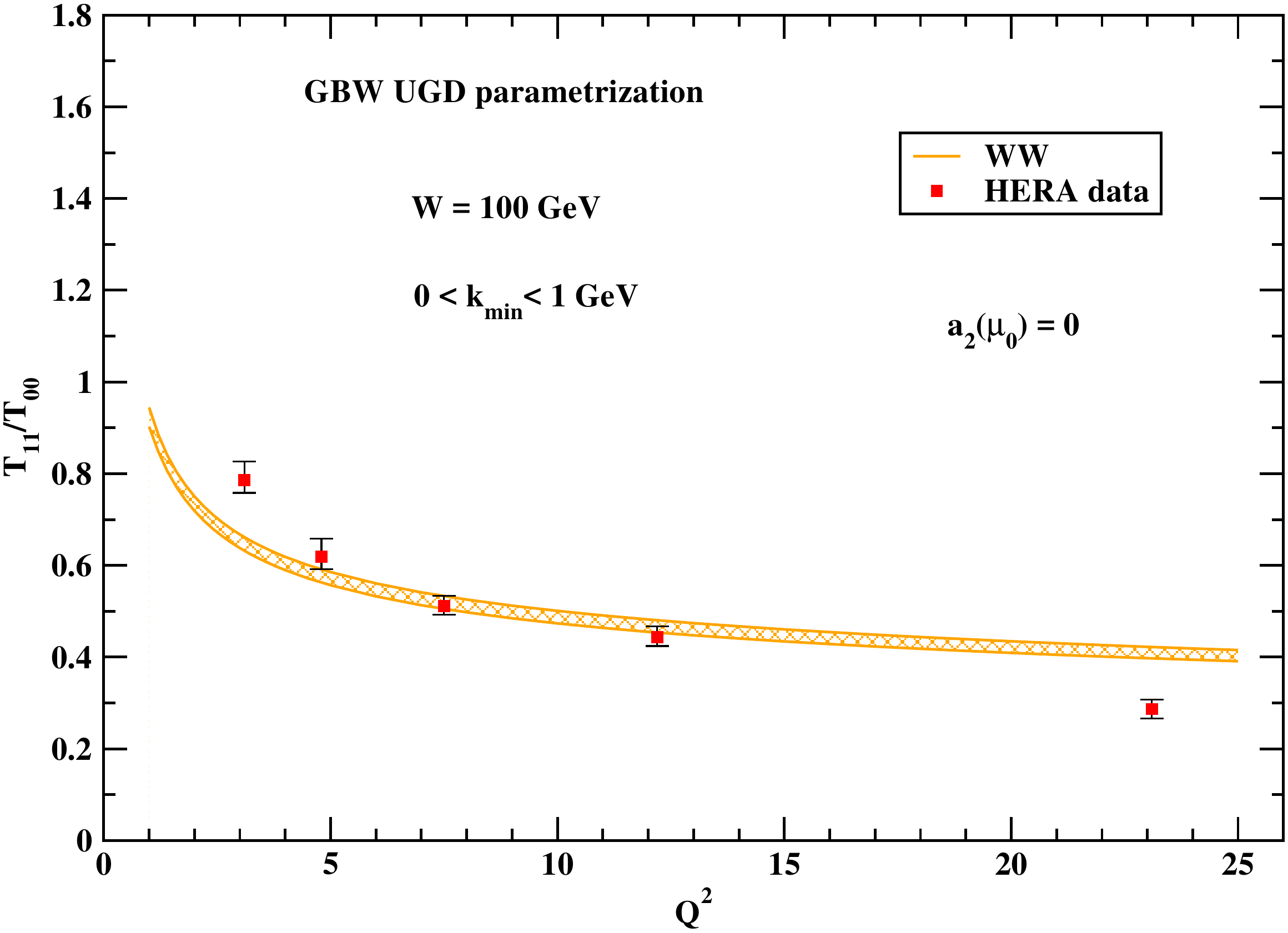}

\caption{$Q^2$-dependence of the helicity-amplitude ratio $T_{11}/T_{00}$ for
  the GBW UGD model at $W = 100$~GeV. The band is the effect of a lower
  cutoff in the $\kappa$-integration, ranging from 0. to 1~GeV. In the
  twist-2 DA we have put $a_2(\mu_0 = 1 \mbox{ GeV}) = 0$ and the $T_{11}$
  amplitude has been calculated in the WW approximation.}
\label{fig:ratio_GBW_kmin}
\end{figure}

The stability of $T_{11}/T_{00}$ under the lower cut-off for $\kappa$, in the
range 0 $< \kappa_{\rm min} < 1$~GeV, has been investigated. This is a
fundamental test since, if passed, it underpins the main underlying
assumption of this work, namely that {\em both} the helicity amplitudes
considered here are dominated by the large $\kappa$ region. In
Fig.~\ref{fig:ratio_GBW_kmin} we show the result of this test for the
GBW model at $W = 100$~GeV; similar plots can be obtained with the other
UGD models, with the only exception of the IN model.
There is a clear indication that the small-$\kappa$ region
gives only a marginal contribution.

\subsection{Tools and systematics}
\label{tools}

All numerical calculations we done in \textsc{Fortran}, making use of specific
CERNLIB routines~\cite{cernlib} to perform numerical integrations and the
computation of (poly-)gamma functions. In order to deal with all the considered
UGD models, we found advantageous to create a modular library which allowed us
to bind and call all sets via a unique and simple interface, serving at the
same time as a working environment for the creation of new, user-customized
UGD parametrizations.

The uncertainty coming from the numerical 2-dimensional over $\kappa$ and $y$
in Eqs.~\eqref{amplitude}, \eqref{Phi_LL} and \eqref{Phi_TT} was directly
estimated by the {\tt Dadmul} integrator~\cite{cernlib} and it was constantly
kept below 0.5\%. In the case of the HSS and WMR models, one should take into
account of an extra-source of systematic uncertainties coming from the
integration on $\nu$ (Eq.~(\ref{HentsUGD})) and on $z$ and $\kappa_t^2$
(Eqs.~(\ref{WMR_UGD}) and~(\ref{WMR_Tg})), respectively. Even in this case,
we managed to keep the numerical error very small.

Furthermore, it is worth to note that for all the UGD models involving the use
of standard PDF parametrizations, it was needed to put a lower cut-off in
$\kappa$, in order to respect the kinematical regime where each set has been
extracted. We gauged the effect of using different PDF parametrizations by making
tests with the most popular sets extracted from global fits, namely
MMHT14~\cite{Harland-Lang:2014zoa}, CT14~\cite{Dulat:2015mca} and
NNPDF3.0~\cite{Ball:2014uwa}, as provided by the LHAPDF Interface
6.2.1~\cite{Buckley:2014ana}, after imposing a provisional cut-off of
$\kappa_{\rm min}^{\rm (test)} = 1$ GeV. We checked that the discrepancy among the
various cases is small or negligible. Then, we did the final calculations by
using the CT14 parametrization, which allowed us to integrate over $\kappa$ down
to $\kappa_{\rm min} = 0.3$ GeV. Results with the gluon momentum derivative and
the WMR model were obtained by imposing such a cut-off, while we adopted a
dynamic strategy as for the IN one: we cut the contribution coming from the
IN hard component (Eq.~(\ref{hardterm})), at $\kappa_{\rm min}$, while no
cut-off was imposed for the soft component  (Eq.~(\ref{softterm})). With
respect to this model, we made further tests by considering the effect of
using different DGLAP inputs (see Table I of Ref.~\cite{Ivanov:2000cm}) for
the parameters $Q_{c}$, $\kappa_{h}$ and $\mu_{\rm soft}$ entering
Eqs.~(\ref{softterm}) and~(\ref{hardterm}). No significant discrepancy among
them was found, so we gave our results for the IN model by using the so-called
CTEQ4L DGLAP input.

Following the definition of IN and WMR models, the PDF set was taken at LO,
while the NLO one was employed in the gluon momentum derivative parametrization.
Moreover, as for the HSS UGD parametrization, we checked that the discrepancy
between the so-called {\em improved} setup given in Eq.~(\ref{ki}) and the
standard one (see, {\it e.g.} Ref.~\cite{Chachamis:2015ona}) is negligible when
considering the helicity-amplitude ratio $T_{11}/T_{00}$.

\section{Discussion}
\label{discussion}

In this paper we have proposed an observable that is well measured in the
experiments at HERA (and could be studied in possible future electron-proton
colliders) -- the dominant helicity amplitudes ratio for the electroproduction
of vector mesons -- as a nontrivial testfield to discriminate the models for
the unintegrated gluon distribution in the proton.

The main motivation of our study are  the features, observed at HERA,  of
polarization observables for exclusive vector meson electroproduction. In the
cases of both longitudinal and transverse polarizations, the measured cross
sections demonstrate similar dependencies on kinematic variables: specific
$Q^2$ scaling, $t$- and $W$-dependencies that are distinct from the ones seen
in soft diffractive exclusive processes.  This  indicates that the dominant
physical mechanism in both cases is the scattering of a small transverse-size,
$\sim 1/Q$, dipole on a proton. 

On the theoretical side we have a description in $\kappa$-factorization,
where the nonperturbative physics is encoded in the unintegrated gluon
distribution, ${\cal F}(x,\kappa^2)$, and in the vector meson twist-2 and twist-3
DAs (which includes both WW and genuine twist-3 contributions), that
parameterize the probability amplitudes for the transition of 2- and 3- parton
 small-transverse-size  colorless states  to the  vector meson.   

In our analysis we  have  considered six models for ${\cal F}(x,\kappa^2)$, which
exhibit rather different shape of $\kappa$-dependence in the  region,
$\kappa^2\sim$ few $\rm{ GeV}^2$, relevant for the kinematic of the $\rho$-meson
electroproduction at HERA, as shown in Fig.~\ref{fig:UGDs_vs_k2}. 

In our numerical study we  have  found rather weak sensitivity of our predictions for
the helicity-amplitude ratio to the physics encoded in the meson DAs (though
values of longitudinal and transverse amplitudes separately depend strongly
on the model for DAs). As an example, in Fig.~\ref{fig:ratio_GBW_evolved} we
 have presented  results for the GBW model of ${\cal F}(x,\kappa^2)$. Here the dominance
of the WW contribution over the genuine twist-3 one is clearly seen. Besides,
we  have  found rather moderate dependence of our observable on the shape of twist-2
DA. Indeed, varying the value of $a_2$ in a wide range         
in comparison to the value $a_2(\mu_0)=0.18 \pm 0.10$ obtained from the QCD sum rules~\cite{Ball:1998sk}, and the one calculated recently on the lattice in Ref.~\cite{Braun:2016wnx}, $a_2(\mu=2 \mbox{ GeV})=0.132 \pm 0.027$, we have found small variation of the amplitude ratio, on the level of the experimental errors.  

Another important issue is the small-size color dipole dominance that allows us
to use results for the $\gamma^*\to \rho$ IF calculated unambiguously in terms
of the meson DAs. To clarify this question we  have introduced a cut-off in the
$\kappa$-integration and studied the stability of our predictions on the
excluded region of small gluon transverse momenta. In
Fig.~\ref{fig:ratio_GBW_kmin}, considering again GBW model as an example, we
 have shown  that the sensitivity of our predictions to the region of small $\kappa$ is
indeed not strong, the variation of our results is lower than or comparable to the
data errors.     

In this way we  have seen  that the dominance of the small-size dipole production
mechanism is supported both by the qualitative features of the data and by the
theoretical calculations in  $\kappa$-factorization.     
This gives evidence to our main statement that, having precise HERA data on the
helicity-amplitude ratio, one can obtain important information about the
$\kappa$-shape of the UGD. To demonstrate this in Fig.~\ref{fig:ratio_all} we
 have confronted  HERA data with the predictions calculated with six different UGD
models. We  have seen  that none of the models is able to reproduce data over the
entire $Q^2$ range and that HERA data on the transverse to longitudinal
amplitudes ratio are really precise enough to discriminate predictions of
different UGD models. 

Our work is closely related to the study of Ref.~\cite{Besse:2013muy}, where the
same process was investigated in much detail in the dipole approach. In this
case the process helicity amplitudes are factorized in terms of the dipole
cross section $\hat \sigma (x,r)$. The  $\kappa$-factorization and the dipole
approach are mathematically related through a Fourier transformation, but the latter approach represents the most natural language to discuss saturation
effects, due to a distinct picture of saturation for the $\hat \sigma (x,r)$
dependence on $r$ for the dipole sizes that exceed the reverse saturation scale,
$r\geq 1/Q_S(x)$. Besides, nonlinear evolution equations that determine the $x$-dependence of $\hat \sigma (x,r)$ and include saturation effects are
formulated in the transverse coordinate space.  

In Ref.~\cite{Besse:2013muy} several models for $\hat \sigma (x,r)$
(including the GBW model adopted by us here) that include saturation
effects, and whose parameters were fitted to describe inclusive DIS data, were
considered. They were used to make predictions for vector meson exclusive
production at HERA kinematics. It was found in Ref.~\cite{Besse:2013muy} that the
predictions of GBW and of other more elaborated dipole models are close to
each other and give rather good, but not excellent, description of HERA data at
virtualities bigger than $Q^2\approx 5\,\rm{GeV}^2$. 

Another interesting issue studied in Ref.~\cite{Besse:2013muy} is the radial
distribution of the dipoles, that contributes to the longitudinal and the
transverse helicity amplitudes for $\rho$-production. It was shown that for
large $Q^2$ both helicity amplitudes are dominated by the contributions of
small size dipoles, which is another source of evidence in favour of
the small-size dipole mechanism for the hard vector meson electroproduction
at HERA.  Besides, as it is shown in Ref.~\cite{Besse:2013muy}, in the case of
large $Q^2$, see the right panels of Fig.~17 in Ref.~\cite{Besse:2013muy} for $Q^2=10\,\rm{GeV}^2$, the  relevant values of $r$ are considerably lower than
those where the dipole cross section $\hat \sigma (x,r)$ starts to saturate.
This is perhaps not surprising, since the estimated value of the saturation
scale at HERA energies is not big, $Q^2_S\sim 1\,\rm{GeV}^2$. Therefore one can
anticipate that saturation effects for hard vector meson electroproduction at
HERA do not play a crucial role. The region of large values of $r$, where
the dipole cross section saturates, represents only a corner of $r$-integration
region for both helicity amplitudes in the dipole approach.\footnote{The
  situation is different in the region of smaller $Q^2$, where the saturation
  region constitutes an essential part of $r$-integration range, see the left
  panels of Fig.~17 in Ref.~\cite{Besse:2013muy}. But in that case one cannot rely
  on the twist expansion in the calculation of the $\gamma^*\to \rho$
  transition, which is expressed in terms of the lowest twist-2 and twist-3
  DAs only. It would be very interesting to consider the same process at
  smaller values of $x$, where the saturation scale is bigger and saturation
  effects are expected to be more pronounced, but this would require
  experiments at larger energies.} In the language of $\kappa$-factorization,
that we use in this work, the saturation region is related to the
$\kappa$-integration region of small $\kappa$. Our calculations for the GBW
model with $\kappa$ cutoff, see Fig.~\ref{fig:ratio_GBW_kmin}, show that,
indeed, the helicity-amplitude ratio for hard meson electroproduction at HERA is not very sensitive to this saturation region. Therefore we believe that
HERA data allow to obtain nontrivial information on the UGD shape (or
equivalently, about  the $r$ shape of the dipole cross section) in the
kinematic range where the {\it linear} evolution regime is still dominant.

Finally, as our closing statement, we recommend that further tests of models
for the unintegrated gluon distribution, as well as possible new model
proposals, take into due account our suggestion to utilize the important
information encoded in the HERA data on the helicity structure in the light
vector meson electroproduction. 

\section{Acknowledgments}

We thank M.~Hentschinski for fruitful discussions. \\
FGC acknowledges support from the Italian Foundation ``Angelo della Riccia''.
AP acknowledges support from the INFN/QFT@colliders project.


\begin{thebibliography}{99}

\bibitem{DGLAP}
V.N.~Gribov, L.N.~Lipatov, Sov. J. Nucl. Phys.  {\bf 15} (1972)  438;
G.~Altarelli, G.~Parisi, Nucl. Phys. B {\bf 126 } (1977) 298;
Y.L.~Dokshitzer, Sov. Phys. JETP {\bf 46 } (1977) 641.

\bibitem{Collins:1996fb}
J.C.~Collins, L.~Frankfurt and M.~Strikman,
Phys.\ Rev.\ D {\bf 56} (1997) 2982
[hep-ph/9611433].
  
\bibitem{Radyushkin:1997ki}
A.V.~Radyushkin,
Phys.\ Rev.\ D {\bf 56} (1997) 5524
[hep-ph/9704207].

\bibitem{BFKL}
V.S.~Fadin, E.A.~Kuraev, L.N.~Lipatov, Phys. Lett. {\bf B60} (1975) 50;
E.A.~Kuraev, L.N.~Lipatov and V.S.~Fadin, Zh. Eksp. Teor. Fiz. {\bf 71} (1976)
840 [Sov. Phys. JETP {\bf 44} (1976) 443]; {\bf 72} (1977) 377 [{\bf 45} (1977)
199];
Ya.Ya.~Balitskii and L.N.~Lipatov, Sov. J. Nucl. Phys. {\bf 28} (1978) 822.

\bibitem{small_x_WG}
J.R.~Andersen {\it et al.} [Small x Collaboration],
Eur. Phys. J. C {\bf 48} (2006) 53 [hep-ph/0604189];
Eur. Phys. J. C {\bf 35} (2004) 67 [hep-ph/0312333];
B.~Andersson {\it et al.} [Small x Collaboration],
Eur. Phys. J. C {\bf 25} (2002) 77 [hep-ph/0204115].

\bibitem{Angeles-Martinez:2015sea}
  R.~Angeles-Martinez {\it et al.},
  Acta Phys.\ Polon.\ B {\bf 46} (2015) no.12,  2501
  [arXiv:1507.05267 [hep-ph]].
  
\bibitem{Aaron:2009xp}
F.D.~Aaron {\it et al.} [H1 Collaboration],
JHEP {\bf 1005} (2010) 032
[arXiv:0910.5831 [hep-ex]].

\bibitem{Chekanov:2007zr}
S.~Chekanov {\it et al.} [ZEUS Collaboration],
PMC Phys.\ A {\bf 1} (2007) 6
[arXiv:0708.1478 [hep-ex]].

\bibitem{Anikin:2009bf}
I.V.~Anikin, D.Yu.~Ivanov, B.~Pire, L.~Szymanowski and S.~Wallon,
Nucl.\ Phys.\ B {\bf 828} (2010) 1
[arXiv:0909.4090 [hep-ph]].

\bibitem{Anikin:2011sa}
I.V.~Anikin, A.~Besse, D.Yu.~Ivanov, B.~Pire, L.~Szymanowski and S.~Wallon,
Phys.\ Rev.\ D {\bf 84} (2011) 054004
[arXiv:1105.1761 [hep-ph]].
  
\bibitem{Besse:2012ia}
A.~Besse, L.~Szymanowski and S.~Wallon,
Nucl.\ Phys.\ B {\bf 867} (2013) 19
[arXiv:1204.2281 [hep-ph]].
  
\bibitem{Besse:2013muy}
A.~Besse, L.~Szymanowski and S.~Wallon,
JHEP {\bf 1311} (2013) 062
[arXiv:1302.1766 [hep-ph]].

\bibitem{Ivanov:1998gk}
D.Yu.~Ivanov, R.~Kirschner,
Phys.\ Rev.\ D {\bf 58} (1998) 114026
[hep-ph/9807324].

\bibitem{Ball:1998sk}
P.~Ball, V.M.~Braun, Y.~Koike and K.~Tanaka,
Nucl.\ Phys.\ B {\bf 529} (1998) 323
[hep-ph/9802299].

\bibitem{Dulat:2015mca}
S.~Dulat et al. 
Phys.\ Rev.\ D {\bf 93} (2016) no.3,  033006
[arXiv:1506.07443 [hep-ph]].

\bibitem{Ivanov:2000cm}
I.P.~Ivanov, N.N.~Nikolaev,
Phys.\ Rev.\ D {\bf 65} (2002) 054004
[hep-ph/0004206].

\bibitem{Nikolaev:1994cd}
N.N.~Nikolaev, B.G.~Zakharov,
Phys.\ Lett.\ B {\bf 332} (1994) 177
[hep-ph/9403281];
Z.\ Phys.\ C {\bf 53} (1992) 331. 

\bibitem{Hentschinski:2012kr}
M.~Hentschinski, A.~Sabio Vera, C.~Salas,
Phys.\ Rev.\ Lett.\  {\bf 110} (2013) no.4,  041601
[arXiv:1209.1353 [hep-ph]].

\bibitem{Chachamis:2015ona}
G.~Chachamis, M.~De\'{a}k, M.~Hentschinski, G.~Rodrigo, A.~Sabio Vera,
JHEP {\bf 1509} (2015) 123
[arXiv:1507.05778 [hep-ph]].

\bibitem{Bautista:2016xnp}
I.~Bautista, A.~Fernandez Tellez, M.~Hentschinski,
Phys.\ Rev.\ D {\bf 94} (2016) no.5,  054002
[arXiv:1607.05203 [hep-ph]].

\bibitem{GolecBiernat:1998js}
K.J.~Golec-Biernat, M.~W\"usthoff,
Phys.\ Rev.\ D {\bf 59} (1998) 014017
[hep-ph/9807513].

\bibitem{Watt:2003mx}
G.~Watt, A.D.~Martin, M.G.~Ryskin,
Eur.\ Phys.\ J.\ C {\bf 31} (2003) 73
[hep-ph/0306169].

\bibitem{cernlib}
CERNLIB Homepage: \url{http://cernlib.web.cern.ch/cernlib}.

\bibitem{Harland-Lang:2014zoa}
L.A.~Harland-Lang, A.D.~Martin, P.~Motylinski, R.S.~Thorne,
Eur.\ Phys.\ J.\ C {\bf 75} (2015) no.5,  204
[arXiv:1412.3989 [hep-ph]].

\bibitem{Ball:2014uwa}
R.D.~Ball {\it et al.} [NNPDF Collaboration],
JHEP {\bf 1504} (2015) 040
[arXiv:1410.8849 [hep-ph]].

\bibitem{Buckley:2014ana}
A.~Buckley, J.~Ferrando, S.~Lloyd, K.~Nordstr{\"o}m, B.~Page, M.~R{\"u}fenacht, M.~Schnherr, G.~Watt,
Eur.\ Phys.\ J.\ C {\bf 75} (2015) 132
[arXiv:1412.7420 [hep-ph]].

\bibitem{Braun:2016wnx}
  V.M.~Braun {\it et al.},
  JHEP {\bf 1704} (2017) 082
  [arXiv:1612.02955 [hep-lat]].
  
\end{thebibliography}
\end{document}